\documentclass[a4paper,12pt, epsfig]{article}
\usepackage{epsfig}
\usepackage{epstopdf}
\usepackage{graphicx}
\usepackage{ifthen}

\usepackage{amsmath}
\usepackage[psamsfonts]{amssymb}
\usepackage{euscript}

\usepackage{latexsym}
\usepackage[arrow,matrix,curve]{xy}

\usepackage{amsmath}
\usepackage[psamsfonts]{amssymb}
\usepackage{euscript}

\usepackage{latexsym}
\usepackage[arrow,matrix,curve]{xy}

\newskip\humongous \humongous=0pt plus 1000pt minus 1000pt

\newif\ifdtup

\def\,{\hspace{-.1cm}}
\def\hsp{,\hspace{.7cm}}

\def\co{{\mathcal{O}}}
\def\df{\mathcal{D}_f}

\def\fc#1#2 {\frac{n}{q}#1\frac{n}{q}#2}

\renewcommand{\theequation}{\arabic{section}.\arabic{equation}}
\renewcommand{\(}{\begin{equation}}
\renewcommand{\)}{end{equation} \vspace{-.05in}\linebreak}

\newcounter{saveeqn}
\newcounter{savealpheqn}

\newcommand{\alpheqn}{\setcounter{saveeqn}{\value{equation}}%
  \stepcounter{saveeqn}\setcounter{equation}{0}%
  \renewcommand{\theequation}{\mbox{\arabic{section}.\arabic{saveeqn}
\alph{equation}}}
  \renewcommand{\)}{\end{equation}}}
\def\part#1{\frac{\partial}{\partial{#1}}}%
\def\group#1{\refstepcounter{equation}\setcounter{saveeqn}
 {\value{equation}}%
  \label{#1}\setcounter{equation}{0}%
\renewcommand{\theequation}{\mbox{\arabic{section}.\arabic{saveeqn}
\alph{equation}}}
  \renewcommand{\)}{\end{equation}}}
\newcommand{\reseteqn}{\setcounter{equation}{\value{saveeqn}}%
  \renewcommand{\theequation}{\arabic{section}.\arabic{equation}}%
  \renewcommand{\)}{\end{equation}}}

\newcommand{\aalpheqn}{\setcounter{saveeqn}{\value{equation}}%
  \stepcounter{saveeqn}\setcounter{equation}{0}%
  \renewcommand{\theequation}{\mbox{
        \Alph{subsection}.\arabic{saveeqn}\alph{equation}}}
   \renewcommand{\)}{\end{equation}}}
\newcommand{\areseteqn}{\setcounter{equation}{\value{saveeqn}}%
  \renewcommand{\theequation}{\Alph{subsection}.\arabic{equation}}%
  \renewcommand{\)}{\end{equation}}}

\renewcommand{\thefootnote}{\alph{footnote}}
\renewcommand{\(}{\begin{equation}}
\renewcommand{\)}{\end{equation}}
\newcommand{\ba}{\begin{eqnarray}}
\newcommand{\ea}{\end{eqnarray}}

\newcommand{\bp}{\mathop{\vtop{\ialign{##\crcr
   $\hfil\displaystyle{}\hfil$\crcr\noalign{\kern-13pt\nointerlineskip}
   \BIG{(}\hskip0pt\crcr\noalign{\kern3pt}}}}}
\newcommand{\cbp}{\mathop{\vtop{\ialign{##\crcr
   $\hfil\displaystyle{}\hfil$\crcr\noalign{\kern-13pt\nointerlineskip}
   \BIG{)}\hskip0pt\crcr\noalign{\kern3pt}}}}}
\newcommand{\pa}{\mathop{\vtop{\ialign{##\crcr
    
$\hfil\displaystyle{\oplus}\hfil$\crcr\noalign{\kern+1pt\nointerlineskip 
}
   \hspace{.08in}$^{\alpha=0}$\hskip6pt\crcr\noalign{\kern3pt}}}}}
\renewcommand{\hsp}{,\hspace{.3in}}

\catcode`\@=11
\def\vereq#1#2{\lower3pt\vbox{\baselineskip1.5pt \lineskip1.5pt
\ialign{$\m@th#1\hfill##\hfil$\crcr#2\crcr\sim\crcr}}}
\catcode`\@=12

\renewcommand{\(}{\begin{equation}}
\renewcommand{\)}{\end{equation}}

\def\exp#1{\hbox{\rm exp}\left(#1\right)}

\newcommand{\beas}{\begin{eqnarray*}}
\newcommand{\eeas}{\end{eqnarray*}}

\newcommand{\bquo}{\begin{quote}}
\newcommand{\enqu}{\end{quote}}

\newcommand{\beq}{\begin{equation}}
\newcommand{\eeq}{\end{equation}}
\newcommand{\bea}{\begin{eqnarray}}
\newcommand{\eea}{\end{eqnarray}}

\newskip\humongous \humongous=0pt plus 1000pt minus 1000pt

\newif\ifdtup

\catcode`\@=11

\@addtoreset{equation}{section}

\def\@normalsize{\@setsize\normalsize{15pt}\xiipt\@xiipt
\abovedisplayskip 14pt plus3pt minus3pt%
\belowdisplayskip \abovedisplayskip
\abovedisplayshortskip \z@ plus3pt%
\belowdisplayshortskip 7pt plus3.5pt minus0pt}

\def\small{\@setsize\small{13.6pt}\xipt\@xipt
\abovedisplayskip 13pt plus3pt minus3pt%
\belowdisplayskip \abovedisplayskip
\abovedisplayshortskip \z@ plus3pt%
\belowdisplayshortskip 7pt plus3.5pt minus0pt
\def\@listi{\parsep 4.5pt plus 2pt minus 1pt
      \itemsep \parsep
      \topsep 9pt plus 3pt minus 3pt}}

\relax

\catcode`@=12

\catcode`\@=11

\def\section{\@startsection{section}{1}{\z@}{3.5ex plus 1ex minus  .2ex}{2.3ex plus .2ex}{\large\bf}}

\def\thesection{\arabic{section}}
\def\thesubsection{\arabic{section}.\arabic{subsection}}

\def\appendix{\setcounter{section}{0}
 \def\thesection{Appendix \Alph{section}}
 \def\thesubsection{\Alph{section}.\arabic{subsection}}
 \def\theequation{\Alph{section}.\arabic{equation}}}
\renewcommand{\theequation}{\arabic{section}.\arabic{equation}}

\begin{document}
\def\thefootnote{\fnsymbol{footnote}}
\def\thetitle{Squeezing the Free Scalar Ground State}
\def\autone{Yao Zhou${}^{1,2}$}
\def\auttwo{Hui Liu${}^{1,2}$}
\def\autthree{Jarah Evslin${}^{1,2}$}
\def\affa{Institute of Modern Physics, Chinese Academy of Sciences, Lanzhou 730000, China}
\def\affb{University of the Chinese Academy of Sciences, Beijing 100049, China}

\begin{center}
{\large {\bf \thetitle}}

\bigskip

\bigskip

{\large \noindent  \autone \footnote{yaozhou@impcas.ac.cn}, \auttwo \footnote{liuhui@impcas.ac.cn} and \autthree \footnote{jarah@impcas.ac.cn} }

\vskip.7cm

1) \affa\\
2) \affb\\

\end{center}

\begin{abstract}
\noindent
Consider two free Hamiltonians for the same scalar field with two different masses.  We find a squeeze operator which maps the ground state of one to the other.  The operator is described in both the Dirac and also the Schrodinger wavefunctional formalisms for quantum field theory.   We conjecture that this construction can be generalized to obtain operators which map between distinct topological sectors in the same theory.

\end{abstract}

%
\setcounter{footnote}{0}
\renewcommand{\thefootnote}{\arabic{footnote}}




\section{Introduction}

In weakly coupled quantum field theories, classical solutions are represented\footnote{However even small quantum corrections can affect the stability of a solution \cite{delfino,davies}.} by operators \cite{hepp,mandelstamkink}.  As the theory flows to strong coupling, the correspondence breaks but in some cases the operator persists while the classical solution loses relevance.  For example, in $\mathcal{N}=2$ super QCD, a large bare mass for the squark hypermultiplets leads to a classical 't Hooft Polyakov monopole.  When this bare mass is tuned down below the strong coupling scale, the BPS monopole remains in the spectrum and if the supersymmetry is softly broken to $\mathcal{N}=1$, it is even responsible for confinement \cite{sw2}.  The monopole which confines is not supported by any Higgs VEV, and so is not a classical solution, it exists only as an operator.  The fact that it condenses results from the fact that it is tachyonic, which presumably can be read from the commutator of the operator and the Hamiltonian.

This motivates us to try to understand how to construct soliton operators explicitly, so that their commutators with the Hamiltonian may be calculated.  Sanity requires that we first attempt this in the weakly coupled regime, before attempting to move to strong coupling.    In Ref.~\cite{mekink} we began the construction of the kink operator in the two-dimensional double well $\phi^4$ theory.   We decomposed the operator $\co$ which constructs the kink as a product of two pieces
\beq
\co=\df \co_1.
\eeq
The first is $\df$, the standard displacement operator which fixes the expectation value of the scalar field to follow, at leading order, the classical solution $f$ as described in Ref.~\cite{taylor78}.  The second piece $\co_1$ was not found explicitly, instead it was found that, at subleading order, it must satisfy 
\beq
b_k\co_1=\co_1 A(a_p)\hsp  b_{BO}\co_1=\co_1 B(a_p)\hsp \pi_0\co_1=\co_1 C(a_p) \label{bog2}
\eeq
where $A$, $B$ and $C$ are arbitrary functions of all of the annihilation operators $a_p$.   Here $a_p$ are the annihilation operators for a free scalar field, while $b_k$, $b_{BO}$ and $\pi_0$ are the analogous operators for another noninteracting quantum field theory, the P\"oschl-Teller theory.   

The relationship between $a_p$ and the $b$ operators is a kind of generalization of a Bogoliubov transform, and so the solution $\co_1$ will be a kind of squeeze operator.  We would like to eventually solve (\ref{bog2}) and so find the operator $\co$ which creates the kink at subleading order.

In the present paper we present what we feel is a necessary first step.  We methodically study the simplest example of a Bogoliubov transform which interpolates between the ground states of two Hamiltonians.  Here, instead of the free Hamiltonian and the P\"oschl-Teller Hamiltonian, we simply consider two free Hamiltonians with two different masses.  The additional complication in the case of the P\"oschl-Teller Hamiltonian will be that the eigenstates are hypergeometric functions instead of plane waves, but we expect that formally the computation will proceed similarly.  As we do not know whether it will be most efficient to proceed using the Dirac or the Schrodinger wavefunctional representation, in this paper we use both.

We begin in Sec.~\ref{qm} by reviewing this standard procedure in quantum mechanics.  Here a squeeze operators maps the ground state of a harmonic oscillator to that of another harmonic oscillator with a different frequency.  The computation in a 1+1 dimensional quantum field theory appears in Sec.~\ref{qft}.

After this paper was posted we became aware of the preprint \cite{cotler} which obtains the same squeeze operator using a new method, which the authors name quantum circuit perturbation theory.

\section{Quantum Harmonic Oscillator} \label{qm}

In quantum mechanics, position eigenstates
\beq
\hat{x}|x\rangle=x|x\rangle
\eeq
provide a basis of the Hilbert space.  Normalizing these, a general state may be written
\beq
\begin{aligned} 
|\psi\rangle=\int dx |x\rangle\langle x| \psi\rangle=\int d x\  \psi(x)|x\rangle\hsp \psi(x)=\langle x|\psi\rangle.
\end{aligned}
\eeq
This one-to-one correspondence between the Dirac ket $|\psi\rangle$ and the Schrodinger wavefunction $\psi(x)$ implies that any computation can be done using either description of quantum states.

In this section, using both the Dirac and Schrodinger representations, we remind the reader that the squeeze operator relates ground states of quantum harmonic oscillators at different frequences.  This is done as a warm-up for the very similar calculation in quantum field theory in Sec.~\ref{qft}.   

\subsection{The Model}

The Hamiltonian of the harmonic oscillator is
\beq
H=\frac{1}{2}\left(\hat{p}^{2}+\omega^{2} \hat{x}^{2}\right),
\eeq
where the frequency $\omega$ is a positive constant and we have set $\hbar=1$ and set the mass  to $m=1$.   $\hat{x}$ and $\hat{p}$ are the position and momentum operators respectively and satisfy the canonical commutation relations $[\hat{x},\hat{p}]=i$.  We work in the Schrodinger picture, where states are defined at a fixed time and operators do not depend on time.  

One can introduce introduce creation and annihilation operators
\beq
\begin{aligned} 
a &=\frac{1}{\sqrt{2}}\left(\omega^{1 / 2} \hat{x}+i \omega^{-1 / 2} \hat{p}\right)\hsp a^{\dagger} &=\frac{1}{\sqrt{2}}\left(\omega^{1 / 2} \hat{x}-i \omega^{-1 / 2} \hat{p}\right)
\end{aligned}
\eeq
which satisfy the Heisenberg algebra $[a,a^\dagger]=1$ and diagonalize the Hamiltonian
\beq
 H=\omega \left( a^{\dagger} a+\frac{1}{2} \right). \label{hdiag}
\eeq
Eq.~(\ref{hdiag}) together with the Heisenberg algebra imply that the ground state $|0\rangle$ satisfies
\beq
a|0\rangle = 0\hsp 
H|0\rangle = E_{0}|0\rangle, \quad E_{0}=\frac{1}{2} \omega
\eeq
while the excited states are obtained by repeatedly operating $a^\dagger$ on $|0\rangle$
\beq
|n\rangle =\frac{1}{\sqrt{n !}}\left(a^{\dagger}\right)^{n}|0\rangle\hsp
H|n\rangle = E_{n}|n\rangle\hsp E_{n}=n \omega+\frac{1}{2} \omega. 
\eeq

In the Schrodinger representation, the canonical commutation relations are equivalent to fixing
\beq
{\hat{p}}=-i\partial_{\hat{x}}
\eeq
and so the eigenvalue equation $H|E\rangle=E|E\rangle$ becomes a wave equation
\beq
H\psi=E\psi\hsp H=\frac{1}{2}\left(-\partial_x^2+\omega^{2} x^{2}\right)
\eeq
whose solutions are
\beq
\begin{aligned} 
\psi_{n}(x)=\langle x|n\rangle=\sqrt{\frac{1}{2^{n} n !}} \left(\frac{\omega}{\pi}\right)^{1/4} \exp{- \frac{\omega x^2}{2}} \cdot H_{n}(\sqrt{\omega} x)
\end{aligned}
\eeq
where the Hermite polynomials are
\beq
\begin{aligned}H_{n}(x)=(-1)^{n} e^{x^{2}} \frac{\mathrm{d}^{n}}{\mathrm{d} x^{n}} e^{-x^2}
\end{aligned}.
\eeq

\subsection{The Squeeze Operator}

In quantum mechanics, the squeeze operator is 
\beq
\begin{aligned}
\hat{S}_r=\exp{\frac{1}{2} r\left(a a- a^\dagger a^\dagger \right)}
\end{aligned}
\eeq
which is easily seen to be unitary.  It acts on annihilation and creation operators via a Bogoliubov transform so that
\beq
a\hat{S}_r=\cosh{(r)}\hat{S}_r a-\sinh(r)\hat{S}_r a^\dagger\hsp
a^\dagger\hat{S}_r=\cosh(r)\hat{S}_r a^\dagger-\sinh(r)\hat{S}_r a.
\eeq
In particular, as $a$ annihilates the ground state,
\beq
a\hat{S}_r|0\rangle=-\sinh(r)\hat{S}_r a^\dagger|0\rangle\hsp
a^\dagger\hat{S}_r|0\rangle=\cosh(r)\hat{S}_r a^\dagger|0\rangle . \label{id}
\eeq

\subsection{Two Frequencies}

For two quantum harmonic oscillators with different frequencies $\omega_1$ and $\omega_2$, the canonical variables have two different decompositions
\beq
\begin{aligned}
\hat{x}&=\frac{1}{\sqrt{2 \omega_1}}\left(a_1+a_1^{\dagger}\right)=\frac{1}{\sqrt{2 \omega_2}}\left(a_2+a_2^{\dagger}\right),\\
\hat{p}&=\frac{1}{i} \sqrt{\frac{\omega_1}{2}}\left(a_1-a_1^{\dagger}\right)=\frac{1}{i} \sqrt{\frac{\omega_2}{2}}\left(a_2-a_2^{\dagger}\right).
\end{aligned}
\eeq
The two sets of operators are related by a Bogoliubov transform
\beq
a_2=ua_1+va_1^\dagger\hsp
a_2^\dagger=ua_1^\dagger+va_1 \label{atoa}
\eeq
where $u$ and $v$ are 
\beq
u=\frac{1}{2}\left(\sqrt{\frac{\omega_2}{\omega_1}}+\sqrt{\frac{\omega_1}{\omega_2}}\right)\hsp
v=\frac{1}{2}\left(\sqrt{\frac{\omega_2}{\omega_1}}-\sqrt{\frac{\omega_1}{\omega_2}}\right).
\eeq

Let $|0\rangle_1$ and $|0\rangle_2$ be the ground states of the two Hamiltonians.  Now using Eqs.~(\ref{id}) and (\ref{atoa}) we may compute
\beq
a_2\hat{S}_r|0\rangle_1=ua_1\hat{S}_r|0\rangle_1+va_1^\dagger\hat{S}_r|0\rangle_1=\hat{S}_r\left(-u\sinh(r)+v\cosh(r)\right)a_1^\dagger|0\rangle_1. \label{cond}
\eeq
If we set
\beq
\tanh(r)=\frac{v}{u}
\eeq
then (\ref{cond}) vanishes, identifying
\beq
|0\rangle_2=\hat{S}_{r}|0\rangle_1. \label{qmprinc}
\eeq 
Thus we have reproduced, in the Dirac representation, the fact that the squeeze operator
\beq
\hat{S}_{r}=\exp{\frac{1}{2}{\rm{arctanh}}\left(\frac{\omega_2-\omega_1}{\omega_2+\omega_1}\right)(a_1^2-a_1^{\dagger2})}
\eeq
maps the ground state of one harmonic oscillator to another.

In the Schrodinger representation, the $n$th eigenstate of the $i$th oscillator has wavefunction proportional to
\beq
(a_i^\dagger)^n\psi_0^{(i)}(x)=\left(\frac{\omega_i}{\pi}\right)^{1/4} \frac{H_n(\sqrt{\omega_i}x)}{(\sqrt{2^n})}\exp{-\frac{\omega_i x^{2}}{2}}
\eeq
where $\psi_0^{(i)}(x)$ is the normalized ground state wavefunction.  The Schrodinger version of (\ref{qmprinc}) is
\beq
\begin{aligned}
\hat{S}_r\psi^{(1)}_0(x)&=\exp {-\frac{1}{2} \hat{a}^{\dagger 2}  \tanh r} \exp{-\frac{1}{2}\left(\hat{a}^{\dagger} \hat{a}+\hat{a} \hat{a}^{\dagger}\right) \ln (\cosh r)}  \exp{\frac{1}{2} \hat{a}^{2} \tanh r}\psi_0^{(1)}\\
&=\frac{\sqrt{2}(\omega_1 \omega_2)^{1/4}}{\sqrt{\omega_1+\omega_2}} \sum_{n=0}^{\infty} \frac{1}{2^{2n} n!}\left(\frac{\omega_1-\omega_2}{\omega_1+\omega_2}\
\right)^{n}H_{2n}(\sqrt{\omega_1}x)\left(\frac{\omega_1}{\pi}\right)^{1/4}\exp {-\frac{\omega_1 x^{2}}{2}}\\
&=\left(\frac{\omega_2}{\pi}\right)^{1/4}\exp{-\frac{\omega_2 x^{2}}{2}}=\psi_0^{(2)}(x)  \label{qmmain}
\end{aligned}
\eeq
where the first equality follows from the Baker-Campbell-Hausdorff relation for SU(1,1), the second uses the energy eigenstate wavefunctions and the last is derived in~\ref{app}.

\section{(1+1)-dimensional Klein-Gordon Theory} \label{qft}

\subsection{The Model}

In quantum field theory, the canonical dynamical variables $\hat{x}$ and $\hat{p}$ are replaced by real scalar fields $\Phi(x)$ and $\Pi(x)$.  The (1+1)-dimensional free scalar field theory is defined by the Hamiltonian
\beq
H=\frac{1}{2}\int dx \left(\Pi^2(x)+\Phi(x)(-\partial_x^2 +m^2)\Phi(x)\right). \label{ham}
\eeq
The field operator $\Phi(x)$ and its conjugate $\Pi(x)$ can be expanded in terms of annihilation and creation operators
\beq
\begin{aligned}
\Phi(x)&=\int\frac{dp}{2\pi}\frac{1}{\sqrt{2\omega(p)}} \left(a(p)+a^{\dagger}(-p)\right)e^{ipx},\\
\Pi(x)&=\int\frac{dp}{2\pi}(-i)\sqrt\frac{\omega(p)}{2}\left(a(p)-a^{\dagger}(-p)\right)e^{ipx} 
\end{aligned}
\eeq
where $\omega(p)=\sqrt{p^2+m^2}$. The operators satisfy the commutation relations
\beq
\left[\Phi(x), \Pi(y)\right]=i \delta(x-y)\hsp
[a(p),a^\dagger(q)]=2\pi\delta(p-q) \label{cr}
\eeq
and $H$ can be diagonalized in terms of $a(p)$ and $a^\dagger(p)$
\beq
\begin{aligned} 
H= \int \frac{dp}{2\pi} \omega(p)\left(a^{\dagger}(p) a(p)+\frac{1}{2}\left[a(p), a^{\dagger}(p)\right]\right) 
\end{aligned}
\eeq 
where the second term is the infinite vacuum energy $E_0$.  It can be removed at one value of $m$ by normal ordering, but cannot be removed at multiple values of $m$ simultaneously.  However it is a scalar operator and so affects only the eigenvalues of the Hamiltonian and not its eigenstates, therefore it will be inconsequential for the construction of ground states of various free Hamiltonians.   As the Hamiltonian is diagonal, the ground state $|0\rangle$ is completely characterized by the condition
\beq
a(p)|0\rangle=0.
\eeq

\subsection{Schrodinger Wavefunctional Representation}

In quantum mechanics one can represent states by either vectors, in Dirac's ket notation, or else Schrodinger's wavefunctions.  Similarly, in quantum field theory states can be represented by kets as above or equivalently by Schrodinger wavefunctionals \cite{stuck,friedrichs}.  In quantum field theory, the field $\Phi(x)$ at each point $x$ is an operator and plays the role played by $\hat{x}$ in quantum mechanics.    Thus, the analogues of the position $\hat{x}$ eigenstates $|x\rangle$ in quantum mechanics are the field $\Phi(x)$ eigenstates $|\varphi\rangle$ in quantum field theory, where $\varphi(x)$ is a real function.  These eigenstates are defined by the eigenvalue equation
\beq
\Phi(x)|\varphi\rangle=\varphi(x)|\varphi\rangle.
\eeq
Note that for each value of $x$, the field $\Phi(x)$ is a distinct operator, and so these states are simultaneous eigenvectors of an infinite number of commuting operators.  

As $\Phi(x)$ is Hermitian, these states form an orthogonal basis of all quantum states, and so any state may be decomposed
\beq
\begin{aligned}
|\Psi\rangle=\int  \mathrm{D}\varphi \Psi[\varphi]|\varphi\rangle \label{swf}
\end{aligned}
\eeq
Here $\mathrm{D}\varphi$ in a measure on the space of functions $\varphi(x)$ and $\Psi[\varphi]$ is the Schrodinger wavefunctional, which plays the role of the wavefunction in quantum mechanics.  Eq.~(\ref{swf}) is a one to one correspondence between Dirac kets $|\Psi\rangle$ and Schrodinger wavefunctionals $\Psi[\varphi]$, and so states can be described equivalently using either formalism.  In the Schrodinger representation, as in quantum mechanics, the action of an operator $\mathcal{O}$ on a wavefunction(al) is defined to be that on the corresponding state in (\ref{swf})
\beq
\mathcal{O}\psi(x)=\langle x|\mathcal{O}|\psi\rangle\Rightarrow
\mathcal{O}\Psi[\varphi]=\langle\varphi|\mathcal{O}|\Psi\rangle.
\eeq
For example
\beq
\hat{x}\psi(x)=x\psi(x)\Rightarrow  \Phi(x)\Psi[\varphi]=\varphi(x) \Psi[\varphi]. \label{phiac}
\eeq

As in quantum mechanics, the canonical commutation relations (\ref{cr}) imply that the canonical momentum may be equivalently expressed as a derivative
\beq
\begin{aligned}
\Pi(x)=-i\frac{\delta}{\delta\Phi(x)}.
\end{aligned}
\eeq
We use the functional derivative notation because it becomes a functional derivative when acting on a state
\beq
\hat{p}\psi(x)=-i\partial_x \psi(x)\Rightarrow  \Pi(x)\Psi[\varphi]=-i\frac{\delta}{\delta \varphi(x)} \Psi[\varphi]. \label{piac}
\eeq

\subsection{Ground State in the Schrodinger Representation}

Now we review the solution of the functional differential equation $H\Psi[\varphi]=E\Psi[\varphi]$. Combining Eqs.~(\ref{ham}), (\ref{phiac}) and (\ref{piac}) this becomes the functional differential equation
\beq
\begin{aligned}
\frac{1}{2} \int dx\left(-\frac{\delta^{2}}{\delta \varphi(x) \delta \varphi(x)}+\varphi(x)(-\partial_x^2+m^2)\varphi(x)\right) \Psi[\varphi]=E \Psi[\varphi] . \label{schqft}
\end{aligned}
\eeq
We are interested in the ground state $\Psi_0(\varphi)$.   Inserting the Ansatz
\beq
\Psi_{0}[\varphi]=\eta \exp {-G[\varphi]}
\eeq
into Eq.~(\ref{schqft}) one obtains
\beq
\begin{aligned}
\frac{1}{2} \int dx\left[ \frac{\delta^2 G[\varphi]}{\delta \varphi^2(x)}-\left(\frac{\delta G[\varphi]}{\delta \varphi(x)}\right)^{2}+\varphi(x)\left(-\partial_x^{2}+m^{2}\right) \varphi(x)\right]=E_{0}(\varphi) \label{anseq}
\end{aligned}
\eeq

We now further refine our Ansatz to
\beq
\begin{aligned}
G[\varphi]=\int dx dy \varphi(x) g(x-y) \varphi(y).
\end{aligned}
\eeq
Inserting this into Eq.~(\ref{anseq}) and matching terms with the same number of powers of $\varphi$, one finds
\beq
\begin{aligned}
\frac{1}{2} \int dx\frac{\delta^2 G[\varphi]}{\delta \varphi^2(x)}=\int dx g(0)=E_0
\end{aligned}
\eeq
and
\beq
\begin{aligned}
\frac{1}{2}\left(\frac{\delta G[\varphi]}{\delta \varphi(x)}\right)^{2}&=2\int dx dy dz\varphi(x)g(x-z)g(z-y)\varphi(y)\\
&=\frac{1}{2}\int dx \varphi(x)\left(-\partial_x^{2}+m^{2}\right) \varphi(x)\\
&=\frac{1}{2} \int dx\varphi(x)\left(-\partial_x^{2}+m^{2}\right) \int dy \delta(y-x)\varphi(y) .
\end{aligned}
\eeq

This second equation will be solved if $g(x-y)$ satisfies
\beq
\begin{aligned}
\int d z g(x-z)g(z-y)=\frac{1}{4}(-\partial_x^2+m^2)\delta(y-x) .
\end{aligned}
\eeq
To solve it, we use the Fourier transform
\beq
\begin{aligned} 
g(x-y)=\int \frac{dp}{2\pi} \tilde{g}(p)e^{i p(x-y)} 
\end{aligned}
\eeq
and find 
\beq
\begin{aligned}
&\tilde{g}^2(p)=\frac{1}{4}(p^2+m^2)\hsp 
\tilde{g}(p)=\frac{1}{2}\sqrt{p^2+m^2}=\frac{1}{2}\omega(p)
\end{aligned}
\eeq

Thus
\beq
\begin{aligned} 
g(x-y)&=\frac{1}{2}\int \frac{d p}{2\pi} \omega(p)e^{i p(x-y)} \label{geq}
\end{aligned}
\eeq
and we recover the same divergent ground state energy as in the Dirac notation
\beq
\begin{aligned}
E _ 0=\int d x g(0)=\frac{1}{2}\int dx\int \frac{d p}{2\pi} \omega(p).
\end{aligned}
\eeq
Finally substituting (\ref{geq}) back into our Ansatz, we reproduce the ground state wavefunction in terms of the Fourier transform $\tilde{\varphi}$
\beq
\begin{aligned}
\Psi_0[\tilde{\varphi}]=\eta\ \exp{-\frac{1}{2}\int\frac{dp}{2\pi}\omega(p)\tilde{\varphi}(p)\tilde{\varphi}(-p)}
\end{aligned}
\eeq
where $\eta$ is an arbitrary scalar.

As a consistency check note that $\Psi_0[\varphi]$ is annihilated by
\beq
\begin{aligned} 
a(p)&=\int d x e^{i p x}\left(\sqrt{\frac{\omega(p)}{2}}\Phi(x)+i \sqrt{\frac{1}{2\omega(p)} }\Pi(x)\right) \\
&=\int d x e^{-i p x}\left(\sqrt{\frac{\omega(p)}{2}} \varphi(x)- \sqrt{\frac{1}{2 \omega(p)}}\frac{\delta}{\delta \varphi(x)}\right) \\ 
&=\sqrt{\frac{\omega(p)}{2}} \tilde\varphi(p)- \sqrt{\frac{1}{2 \omega(p)}}(2\pi)\frac{\delta}{\delta \tilde\varphi(-p)}
\end{aligned}
\eeq
as

\beq
\begin{aligned}
&\varphi(x) =\int \frac{dp}{2 \pi} \tilde{\varphi}(p) e^{i px}, \quad \frac{\delta}{\delta \varphi(x)} =\int dp \frac{\delta}{\delta \tilde{\varphi}(p)} e^{-i px} \\
&\frac{\delta \Psi_{0}[\tilde\varphi]}{\delta \tilde\varphi(p)}=-\frac{\omega(p)}{2\pi}\tilde\varphi(-p)\Psi_{0}[\tilde\varphi] .
\end{aligned}
\eeq

\subsection{Normalization}

As our operator which maps one ground state to another will be unitary, it will also preserve the normalization of the wavefunctional.  This normalization can be defined by standard methods if one compactifies the spatial dimension.  In this case the measure $\mathrm{D}\varphi$ is the product of the ordinary measure for the infinitely many Fourier components (each divided by 2$\pi$) of the function $\varphi(x)$ and an inner product can be defined by integration
\beq
\begin{aligned}
\langle\psi | \phi\rangle &=\int \mathrm{D} \varphi\langle\psi | \varphi\rangle\langle \varphi| \phi\rangle=\int \mathrm{D} \varphi \psi^{*}[\varphi] \phi[\varphi].  
\end{aligned}
\eeq
Then
\beq
\begin{aligned} 
1&=\int\mathrm{D}\varphi\ \Psi^*[\varphi]\Psi[\varphi]=\eta^2 \int \mathrm{D}\tilde{\varphi}\ \exp{-\int \frac{d p}{2\pi}\omega(p)\tilde{\varphi}(p)\tilde{\varphi}(-p)} \\
&=\eta^2\mathrm{Det}\left(\sqrt\frac{\pi}{\omega}\right)\hsp \omega_{pq}=\omega(p)\delta_{pq} .
\end{aligned}
\eeq
Thus the wavefunctional
\beq
\begin{aligned} 
\Psi_{0}[\tilde{\varphi}] &=\mathrm{Det}\left(\frac{\omega}{\pi}\right)^{\frac{1}{4}} \exp {-\frac{1}{2} \int \frac{d p}{2\pi}\omega(p) \tilde{\varphi}(p) \tilde{\varphi}(-p)}
\end{aligned}
\eeq
has unit normalization.

Below we will also need certain excited states.  Using the substitution\footnote{This substitution changes the normalization of $\phi$ and so also $a$ and $a^\dag$.}
\beq
\begin{aligned}
&\mathrm{Det}\left(\frac{\omega}{\pi}\right)^{\frac{1}{4}} \longrightarrow \prod_p \left(\frac{\omega(p)}{\pi}\right)^{\frac{1}{4}},\\ &\exp {-\frac{1}{2} \int \frac{d p}{2\pi}\omega(p) \tilde{\varphi}(p) \tilde{\varphi}(-p)} \longrightarrow \prod_p \exp {-\frac{1}{2} \frac{\omega(p)}{2\pi} \tilde{\varphi}(p) \tilde{\varphi}(-p)},
\end{aligned}
\eeq
the wavefunctional can be rewritten as
\beq
\begin{aligned}
\Psi_{0}[\tilde{\varphi}] = \prod_{p}\left(\frac{\omega(p)}{\pi}\right)^{\frac{1}{4}} \exp {-\frac{1}{2} \frac{\omega(p)}{2 \pi}  \tilde\varphi(p)\tilde\varphi(-p)} \label{prodrep}
\end{aligned}
\eeq
which, as expected, is just the infinite product of harmonic oscillator wavefunctions.  $a^\dagger(p)$ excites the mode $p$.  So the functional derivative becomes an ordinary derivative and the $\delta$-function changes from Dirac to Kronecker
\beq
\begin{aligned} a^\dagger(p) \Psi_0[\tilde\varphi] &=\left(\sqrt{\frac{\omega(p)}{2}} \tilde\varphi(p)- \sqrt{\frac{1}{2 \omega(p)}}(2\pi)\frac{d}{d\tilde\varphi(-p)}\right)\left(\frac{\omega(p)}{\pi}\right)^{\frac{1}{4}} \exp {-\frac{1}{2} \frac{ \omega(p)}{2 \pi} \tilde\varphi(p)\tilde\varphi(-p)} \\ 
&\times \prod_{k \neq p}\left(\frac{\omega(k)}{\pi}\right)^{\frac{1}{4}} \exp {-\frac{1}{2} \frac{\omega(k)}{2 \pi}  \tilde\varphi(k)\tilde\varphi(-k)} \\
&=\sqrt{2\omega(p)}\tilde\varphi(p) \prod_{k}\left(\frac{\omega(k)}{\pi}\right)^{\frac{1}{4}} \exp {-\frac{1}{2} \frac{\omega(k)}{2 \pi}  \tilde\varphi(k)\tilde\varphi(-k)}.
\end{aligned}
\eeq
 
With this normalization, $[a^\dagger(p),a(p)]=2\pi$ in the product representation of $\Psi_0[\tilde\varphi]$ in Eq.~(\ref{prodrep}).  Let us define
\beq
a_p\equiv\frac{1}{\sqrt{2\pi}}a(p)=\frac{1}{\sqrt{2}}\left(\sqrt{\frac{\omega(p)}{2\pi}} \tilde\varphi(p)+ \sqrt{\frac{2\pi}{ \omega(p)}}\frac{d}{d\tilde\varphi(-p)}\right).
\eeq
Then the commutor is $[a_p,a_p^\dagger]=1$ and we have
\beq
\begin{aligned}
\left(a^{\dagger}_p a^{\dagger}_{-p}\right)^{n}\Psi_{0}[\varphi]=\frac{ H_{2n}\left(\frac{\omega(p)}{2\pi}\tilde\varphi(p)\tilde\varphi(-p)\right)}{2^n} \Psi_0[\varphi]
\end{aligned}
\eeq
where
\beq
\begin{aligned}
H_{2 n}\left(A(p) \tilde{\varphi}(p) \tilde{\varphi}(-p)\right)=e^{ A(q)\tilde{\varphi}(q) \tilde{\varphi}(-q)} \frac{d^{2 n}}{\left(A(p)d \tilde{\varphi}(p) d \tilde{\varphi}(-p) \right)^{n}} e^{- A(q) \tilde{\varphi}(q) \tilde{\varphi}(-q)}. 
\end{aligned}
\eeq
$A(q)$ is any even function of $q$.

\subsection{The Map Between Ground States}

Now consider two masses $m_1$ and $m_2$.  To be more precise, we are considering a single real scalar field $\Phi(x)$, and a single Hilbert space of states.  However two different operators can be constructed which act on this Hilbert space, which would be interpreted as the Hamiltonians of two distinct theories, one a free scalar with mass $m_1$ and another a free scalar with mass $m_2$.  Whether either of these operators really is the Hamiltonian of the theory will be irrelevant for our discussion.  The conjugate momentum field $\Pi(x)$ is defined not using the Hamiltonian or Lagrangian, but simply by defining it to be the operator which satisfies the canonical commutation relations with $\Phi(x)$.  This is sufficient to define an algebra of operators and the states on which they act.

For each mass, there is a decomposition of the fields
\beq
\begin{aligned}
\int d x\Phi(x)e^{-i p x}&=\frac{1}{\sqrt{2\omega_1(p)}}\left(a_1(p)+a_1^\dagger(-p)\right)=\frac{1}{\sqrt{2\omega_2(p)}}\left(a_2(p)+a_2^\dagger(-p)\right),\\
\int d x\Pi(x)e^{-ipx}&=(-i)\sqrt{\frac{\omega_1(p)}{2}}\left(a_1(p)-a_1^\dagger(-p)\right)=(-i)\sqrt{\frac{\omega_2(p)}{2}}\left(a_2(p)-a_2^\dagger(-p)\right)
\end{aligned}
\eeq
where $\omega_i(p)\equiv\sqrt{p^2+m_i^2}$.  These two decompositions are related by a Bogoliubov transform, similar to (\ref{atoa}) 
\beq
\begin{aligned}
a_2(p)&=u(p)a_1(p)+v(p)a_1^\dagger(-p)\\
u(p)&=\frac{1}{2}\left(\sqrt{\frac{\omega_2(p)}{\omega_1(p)}}+\sqrt{\frac{\omega_1(p)}{\omega_2(p)}}\right)\hsp
v(p)=\frac{1}{2}\left(\sqrt{\frac{\omega_2(p)}{\omega_1(p)}}-\sqrt{\frac{\omega_1(p)}{\omega_2(p)}}\right). \label{bogqft}
\end{aligned}
\eeq

So far everything resembles the quantum mechanics case, albeit with an additional dependence on $p$.  This motivates an Ansatz for our squeeze operator
\beq
\hat{S}=\exp{A}\hsp A=\frac{1}{2}\int\frac{dp}{2\pi} f(p)\left(a_1(p)a_1(-p)- a_1^\dagger(p)a_1^\dagger(-p) \right) \label{sanzatz}
\eeq
for an unknown function $f(p)$.   Note that the operator will be unitary for any real function $f(p)$.  The operator is not normal-ordered, as this would ruin the unitarity\footnote{Construction of a nonunitary, normal-ordered operator is quite easy, one can simply omit the $a^2$ term and repeat the calculation below.  The calculation is much simpler in the nonunitary case as commutators with $a^\dag$ vanish, yielding a constant $f(p)$.}.

 Exponentiating the commutators
\beq
\left[A,a_1(p)\right]= f(p)a_1^\dagger(-p)\hsp
[A,a_1^\dagger(-p)]= f(p) a_1(p), \label{scom}
\eeq
one arrives at
\bea
\hat{S}^\dagger a_1(p)\hat{S}&=&e^{-A} a_1(p) e^{A}=a_1(p)+[-A,a_1(p)]+\frac{1}{2}[-A,[-A,a_1(p)]]+\cdots\nonumber\\
&=&\sum_{n=0}^{\infty} \frac{f(p)^{2 n}}{(2 n) !}a_1(p) - \sum_{n=0}^{\infty} \frac{f(p)^{2 n+1}}{(2 n+1) !}a_1^{\dagger}(-p)\nonumber\\
&=&\cosh(f(p))a_1(p)-\sinh(f(p))a_1^\dagger(-p) \nonumber \\
\hat{S}^\dagger a^\dagger_1(-p)\hat{S}&=&    \cosh(f(p))a_1^\dagger(-p)-\sinh(f(p))a_1(p). \label{exp}
\eea

Applying the second annihilation operator at any $p$ to the squeezed state one then finds
\beq
a_2(p)\hat{S}|0\rangle_1=\left(u(p)a_1(p)+v(p)a_1^\dagger(-p)\right)\hat{S}|0\rangle_1=\hat{S}\left(u(p)\cosh(f(p))-v(p)\sinh(f(p))\right)a_1^\dagger|0\rangle_1
\eeq
which vanishes if
\beq
f(p)={\rm{arctanh}}\left(\frac{v(p)}{u(p)}\right)
\eeq
identifying the squeezed state as $|0\rangle_2$.   In summary, we have found that the operator
\beq
\hat{S}=\exp {\frac{1}{2}\int\frac{dp}{2\pi}{\rm{arctanh}}\left(\frac{v(p)}{u(p)}\right)\left(a_1(p)a_1(-p)- a_1^\dagger(p)a_1^\dagger(-p) \right)} \label{sfin}
\eeq
maps the ground state of the free scalar Hamiltonian with mass $m_1$ to that with mass $m_2$
\beq
\hat{S}|0\rangle_1=|0\rangle_2.
\eeq

Also, we can check this in the Schrodinger representation
\beq
\begin{aligned} 
\hat{S} \Psi_{0}^{(1)}[\tilde{\varphi}]=&\left\{\prod_{p} \exp {-\frac{1}{2} a_{p}^{\dagger} a_{-p}^{\dagger} \tanh f(p)} \exp {-\frac{1}{2}\left(a_{p}^{\dagger} a_{p}+a_{-p} a_{-p}^{\dagger}\right) \ln (\cosh f(p))}\right.\\
&\left. \times \exp {\frac{1}{2} a_{p} a_{-p} \tanh f(p)}\right\} \Psi_{0}^{(1)}[\tilde{\varphi}] \\
=&\left(\prod_{p} \frac{\sqrt{2}\left(\omega_{1}(p) \omega_{2}(p)\right)^{1 / 4}}{\sqrt{\omega_{1}(p)+\omega_{2}(p)}} \sum_{n=0}^{\infty} \frac{1}{2^{2 n} n !}\left(\frac{\omega_{1}(p)-\omega_{2}(p)}{\omega_{1}(p)+\omega_{2}(p)}\right)^{n} H_{2 n}\left(\frac{\omega_{1}(p)}{2 \pi} \tilde{\varphi}(p) \tilde{\varphi}(-p)\right)\right) \Psi_{0}^{(1)}[\tilde{\varphi}] \\
=&\Psi_{0}^{(2)}[\tilde{\varphi}] \label{sqft}
\end{aligned}
\eeq
where
\beq
\begin{aligned} 
\Psi^{(i)}_{0}[\tilde{\varphi}] = \prod_{p}\left(\frac{\omega_i(p)}{\pi}\right)^{\frac{1}{4}} \exp {-\frac{1}{2} \frac{\omega_i(p)}{2 \pi}  \tilde\varphi(p)\tilde\varphi(-p)}.
\end{aligned}
\eeq
This calculation is similar to that in quantum mechanics.  The last line can be checked by expanding $\Psi^{(i)}_{0}[\tilde{\varphi}]$ at every fixed $q$ in $(\tilde\varphi(q)\tilde\varphi(-q))^l$, as is shown in~\ref{apq}.

\section{Remarks}

We set out to tailor an operator which performs the Bogoliubov transformation (\ref{bogqft}).  In quantum field theory there are an infinite number of $a(p)$ and $a^\dagger(p)$, however this transformation only mixes individual pairs and so it is essentially the same as the quantum mechanical case.  Therefore the squeeze operator (\ref{sfin}) which does the transform is essentially a product of quantum mechanical squeeze operators.  The critical simplification can be seen in the commutation relation (\ref{scom}) which, since it only interchanges two elements, is easily exponentiated to (\ref{exp}).

This operator is similar to that found by Fan and Fan in \cite{hongyi2000} in 3+1 dimensions.  They argue that their operator rescales scalar fields.  In 3+1 dimensions, scalar fields are dimensionful, and so such a scaling corresponds to a dilation.  Similarly, in our case rescaling the mass is equivalent to a dilation.  

In Eq.~(\ref{bog2}) on the other hand the two sets of oscillators, corresponding to fluctuations about the trivial and the one-kink sector respectively, mix all of the $a(p)$ and $a^\dagger(p)$ with no simple factorization into pairs.  The transformation, at the subleading order considered in Ref.~\cite{mekink}, is still linear and so we believe that an Ansatz similar to (\ref{sanzatz}) can still be used.  However, momentum is no longer conserved as the kink breaks translation invariance and so the Ansatz should be generalized to
\beq
\hat{S}=\exp{A}\hsp A=\frac{1}{2}\int\frac{dp}{2\pi} \int\frac{dq}{2\pi} f(p,q)\left(a(p)a(-q)- a^\dagger(p)a^\dagger(-q) \right). \label{newans}
\eeq
Now each commutator in (\ref{exp}) will have an additional convolution with $f$ at each order.  Therefore our final equation for $f$ will involve an infinite series of convolutions.  However it will be a single, algebraic equation for $f$ and so we believe that it can be solved numerically to yield the squeeze operator.   

If one does not demand that the squeeze operator be unitary, one can do much better.  In that case one can keep only the $a^\dagger a^\dagger$ terms in (\ref{newans}).  Then $\hat{S}$ will commute with $a^\dagger$ and only a single commutator with $a$ will be nonvanishing, as after one commutator the $a$ becomes an $a^\dagger$ and so commutes with all other terms in $S$.  This makes the exponentiation in (\ref{exp}) trivial and so one can easily find $f(p,q)$ analytically for any transformation between free theories.  The normalization can be chosen that the resulting ground state is normalized if desired.  However, being nonunitary, of course the squeeze operator will not preserve the normalization of any other states on which it acts.

Needless to say, we believe that the squeeze operator which maps the ground state of any linearized sector of a scalar field theory to the ground state of any other can be found similarly.

\appendix

\section{Squeezing Wavefunctions}\label{app}
In this appendix we will show that $\hat{S}_r \psi_0^{(1)}(x)=\psi_0^{(2)}(x)$.  Dividing Eq.~(\ref{qmmain}) by a constant, this condition is 
\beq
\begin{aligned}
\sum_{n=0}^{\infty} \frac{1}{2^{2 n} n !}\left(\frac{\omega_{1}-\omega_{2}}{\omega_{1}+\omega_{2}}\right)^{n} H_{2 n}(\sqrt{\omega_{1}} x) = \sqrt{\frac{\omega_{1}+\omega_{2}}{2 \omega_{1}}} \exp {\frac{\omega_1-\omega_{2} }{2}x^{2}}. \label{eqa}
\end{aligned}
\eeq
Using the expansion
\beq
\begin{aligned}
H_{2n}(\sqrt{\omega_1}x) =(2n)!\sum_{l=0}^n\frac{(-1)^{n-l}}{(2l)!(n-l)!}(2\sqrt{\omega_1}x)^{2l}
\end{aligned}
\eeq
we can expand the left-hand side of (\ref{eqa}) as
\beq
\begin{aligned}
LHS =\sum_{n=0}^{\infty} \frac{(2n)!}{2^{2 n} n !}\left(\frac{\omega_1 - \omega_2}{\omega_1 + \omega_2}\right)^{n}\sum_{l=0}^n\frac{(-1)^{n-l}2^{2l}\omega_1^l}{(2l)!(n-l)!}x^{2l}
=\sum_{l=0}^{\infty} A_l x^{2l}  \\
\end{aligned}
\eeq
where the coefficients $A_l$ are
\beq
\begin{aligned}
A_l &=\sum_{n=l}^{\infty} \frac{(2n)!}{2^{2 n} n !}\left(\frac{\omega_1 - \omega_2}{\omega_1 + \omega_2}\right)^{n}\frac{(-1)^{n-l}2^{2l}\omega_1^l}{(2l)!(n-l)!} \\
&=\sum_{n=0}^{\infty} \frac{(2(n+l)!}{2^{2 n} (n+l) !}\left(\frac{\omega_1 - \omega_2}{\omega_1 + \omega_2}\right)^{n+l} \frac{(-1)^{n}\omega_1^l}{(2l)!n!} .
\end{aligned}
\eeq

Similarly we may expand the right hand side of (\ref{eqa}) as
\beq
RHS=\sum_{l=0}^{\infty} B_l x^{2l}\hsp
B_l=\frac{1}{l!}\sqrt{\frac{\omega_1+\omega_2}{\omega_1}}\left(\frac{\omega_1-\omega_2}{2}\right)^l. 
\eeq

Before we may compare $A_l$ and $B_l$, we will need some identities.  First
\beq
\frac{(2m)!}{m!}=2^{2m}\left(m-\frac{1}{2}\right)\left(m-\frac{3}{2}\right)\cdots\left(\frac{1}{2}\right). \label{meq}
\eeq
Eq.~(\ref{meq}) at $m=n+l$ divided by (\ref{meq}) at $m=l$ yields the identity
\beq
\begin{aligned} 
 \frac{(2(n+l))!}{(n+l)!}&=\frac{(2l)!}{l!} 2^{2n}\left(n+l-\frac{1}{2}\right)\left(n+l-\frac{3}{2}\right)\cdots\left(l+\frac{1}{2}\right).\\\label{eq1}
\end{aligned}
\eeq
We will also need the Taylor expansion
\beq
\begin{aligned} 
(1+x)^{-k-\frac{1}{2}}&=\sum_{n=0}^{\infty}\frac{1}{n!}\left(-k-\frac{1}{2}\right)\left(-k-\frac{3}{2}\right) \cdots\left(-k-n+\frac{1}{2}\right)x^{n}\\
&=\sum_{n=0}^{\infty}\frac{(-1)^n}{n!}\left(k+\frac{1}{2}\right)\left(k+\frac{3}{2}\right) \cdots\left(k+n-\frac{1}{2}\right)x^{n}. \label{eq2}
\end{aligned}
\eeq

Finally we are ready to compute
\beq
\begin{aligned}
A_l&=\sum_{n=0}^{\infty}\left(n+l-\frac{1}{2}\right)\left(n+l-\frac{3}{2}\right)\cdots\left(l+\frac{1}{2}\right) \left(\frac{\omega_1-\omega_2}{\omega_1+\omega_2}\right)^{n+l} \frac{(-1)^{n}\omega_1^l}{n!l!}\\
&=\left(\frac{\omega_1-\omega_2}{\omega_1+\omega_2}\right)^l\left(1+\left(\frac{\omega_1-\omega_2}{\omega_1+\omega_2}\right)\right)^{-l-\frac{1}{2}}\frac{\omega_1^l}{l!} \\
&=\frac{1}{l!}\sqrt{\frac{\omega_1+\omega_2}{\omega_1}}\left(\frac{\omega_1-\omega_2}{2}\right)^l =B_l 
\end{aligned}
\eeq
where the first equality comes from (\ref{eq1}) and the second from (\ref{eq2}).  Thus we have proved (\ref{eqa}) at every order $2l$.

\section{Squeezing Wavefunctionals}\label{apq}
In this appendix we will show that $\hat{S}_r \Psi_0^{(1)}[\tilde{\varphi}]=\Psi_0^{(2)}[\tilde{\varphi}]$.  Dividing Eq.~(\ref{sqft}) through by a constant factor, this is equivalent to
\bea
\prod_p\alpha_p&=&\prod_p \beta_p\\
\alpha_p&=&\sum_{n=0}^{\infty}  \frac{1}{2^{2 n} n !}\left(\frac{\omega_{1}(p)-\omega_{2}(p)}{\omega_{1}(p)+\omega_{2}(p)}\right)^{n} H_{2 n}\left(\frac{\omega(p)}{2\pi} \tilde{\varphi}(p)\tilde{\varphi}(-p)\right)\nonumber \\
\beta_p&=&\left(\sqrt{\frac{\omega_{1}(p)+\omega_{2}(p)}{2 \omega_{1}(p)}}\right)  \exp {\frac{1}{2} \frac{\omega_2(p)-\omega_1(p)}{2\pi}\tilde{\varphi}(p) \tilde{\varphi}(-p)}.\nonumber
\eea
The coefficients $\alpha_p$ and $\beta_p$ may then be expanded in powers of $(\tilde{\varphi}(p) \tilde{\varphi}(-p))$
\beq
\alpha_p=\prod_{l=0}^{\infty}A_{p,l}\hsp
\beta_p=\prod_{l=0}^{\infty}B_{p,l}
\eeq
where 
\beq
\begin{aligned} A_{p,l} &=\sum_{n=l}^{\infty} \frac{(2 n) !}{2^{2 n} n !}\left(\frac{\omega_{1}(p)-\omega_{2}(p)}{\omega_{1}(p)+\omega_{2}(p)}\right)^{n} \frac{(-1)^{n-l} 2^{2 l} \omega_{1}(p)^{l}}{(2 l) !(n-l) ! (2\pi)^l} \\ 
&=\sum_{n=0}^{\infty} \frac{(2(n+l) !}{2^{2 n}(n+l) !}\left(\frac{\omega_{1}(p)-\omega_{2}(p)}{\omega_{1}(p)+\omega_{2}(p)}\right)^{n+l} \frac{(-1)^{n} \omega_{1}(p)^{l}}{(2 l) ! n !(2\pi)^l}\\ 
B_{p,l}&=\frac{1}{l !} \sqrt{\frac{\omega_{1}(p)+\omega_{2}(p)}{\omega_{1}(p)}}\left(\frac{\omega_{1}(p)-\omega_{2}(p)}{2}\right)^{l}\frac{1}{(2\pi)^l}.
\end{aligned}
\eeq
These are just the $A_l$ and $B_l$ from~\ref{app}.   Therefore $A_{p,l}=B_{p,l}$ and so $\alpha_p=\beta_p$ and (\ref{sqft}) follows.

\section* {Acknowledgement}

\noindent
JE is supported by the CAS Key Research Program of Frontier Sciences grant QYZDY-SSW-SLH006 and the NSFC MianShang grants 11875296 and 11675223.  JE also thanks the Recruitment Program of High-end Foreign Experts for support.


\end{document}